\documentstyle[prl,aps]{revtex}
\input epsf

\tighten
\begin{document}

\def\be{\begin{equation}}
\def\ee{\end{equation}}
\def\ba{\begin{eqnarray}}
\def\ea{\end{eqnarray}}
\def\bq{\begin{quote}}
\def\eq{\end{quote}}
\def\PL{{ \it Phys. Lett.} }
\def\PRL{{\it Phys. Rev. Lett.} }
\def\NP{{\it Nucl. Phys.} }
\def\PR{{\it Phys. Rev.} }
\def\MPL{{\it Mod. Phys. Lett.} }
\def\IJMP{{\it Int. J. Mod .Phys.} }
\newcommand{\labell}[1]{\label{#1}\qquad_{#1}} 
\newcommand{\labels}[1]{\vskip-2ex$_{#1}$\label{#1}} 
\newcommand\gapp{\mathrel{\raise.3ex\hbox{$>$}\mkern-14mu
\lower0.6ex\hbox{$\sim$}}}
\newcommand\gsim{\gapp}
\newcommand\gtsim{\gapp}
\newcommand\lapp{\mathrel{\raise.3ex\hbox{$<$}\mkern-14mu
\lower0.6ex\hbox{$\sim$}}}
\newcommand\lsim{\lapp}
\newcommand\ltsim{\lapp}
\newcommand\M{{\cal M}}
\newcommand\order{{\cal O}}

\newcommand\extra{{\rm {extra}}}
\newcommand\FRW{{\rm {FRW}}}
\newcommand\brm{{\rm {b}}}
\newcommand\ord{{\rm {ord}}}
\newcommand\Pl{{\rm {pl}}}
\newcommand\Mpl{M_{\rm {pl}}}
\newcommand\mgap{m_{\rm {gap}}}
\newcommand\gB{g^{\left(\rm \small B\right)}}

\preprint{SU-GP-01/5-1\\ May 2001}

\draft
\title{Homogeneity, Flatness and ``Large"
Extra Dimensions}

\author{Glenn D. Starkman$^1$, Dejan Stojkovic$^1$ and
Mark Trodden$^2$
}

\address
{
$^1$ Department of Physics \\
Case Western Reserve University \\
Cleveland, OH 44106-7079, USA \\
\vspace*{0.2in}
$^2$ Department of Physics \\
Syracuse University \\
Syracuse, NY, 13244-1130 USA \\
}

\wideabs{

\maketitle

\begin{abstract}
\widetext We consider a
model in which the universe is the direct product of a
(3+1)-dimensional Friedmann, Robertson-Walker (FRW) space and a
compact hyperbolic manifold (CHM). Standard Model fields are
confined to a point in the CHM (i.e. to a brane). In such a
space, the decay of massive Kaluza-Klein modes leads to the
injection of any initial bulk entropy into the observable (FRW)
universe. Both Kolmogoro-Sinai mixing due to the
non-integrability of flows on CHMs and the large statistical
averaging inherent in the collapse of the initial entropy onto
the brane smooth out any initial inhomogeneities in the
distribution of matter and of 3-curvature on any slice of
constant 3-position. If, as we assume, the initial densities and
curvatures in each fundamental correlation volume are drawn from
some universal underlying distributions independent of location
within the space, then these smoothing mechanisms effectively
reduce the density and curvature inhomogeneities projected onto
the FRW. This smoothing is sufficient to account for the current
homogeneity and flatness of the universe. The fundamental scale 
of physics can be $\lsim 1$TeV.  All relevant mass and length scales can have 
natural values in fundamental units.  All large dimensionless 
numbers, such as the entropy of the universe, are understood as 
consequences of the topology of spacetime which is not explained. No model for
the origin of structure is proffered.
\end{abstract}
\pacs{PACS:12.10.-g, 11.10.Kk,
\hfill  SU-GP-01/5-1\\
11.25.M,04.50.+h
\hfill }
}

\narrowtext
The  notion 
\cite{Anton,ADD,RS,otherearly}
that space has more than three space and one time dimensions,
and  thus that our $3\!+\!1$ dimensional universe is merely a
subspace (``the brane") on which ordinary Standard Model (SM) fields are
confined inside a higher dimensional space (``the bulk") in which
gravitational degrees of freedom propagate,
offers new perspectives for cosmology \cite{cp,ekpy}. 
A simple realization is a space-time 
which factorizes into 
an ordinary
dynamical (3+1)-dimensional 
Friedmann-Robertson-Walker universe $\M_{FRW}$ 
and a static extra-dimensional spatial manifold, $\M_\extra$; 
so that $\M=\M_{FRW} \times \M_\extra$.
The brane on which standard model fields are localized
is taken to be stationary at a point  in $\M_\extra$. Previously, we
showed \cite{cp} that if $\M_{extra}$ has hyperbolic, rather than
flat, geometry, then the dynamics of the very early universe
alleviate some standard cosmological problems.
In this letter we argue that the homogeneity and
local flatness of the universe are consequences of a
broad class of initial conditions for such a universe, assuming
that $\M_{extra}$ is sufficiently topologically complex and
that the fundamental laws of physics are universal.

The principal motivation for large extra dimensions was to solve
the gauge hierarchy problem. For an extra dimensional manifold of volume
$V_\extra$,
the force due to massless gravitons will,
by Gauss's law,
follow an inverse square law with an
effective coupling of
\be
\label{Vextra} \Mpl^{-2} = M_{F}^{-(d+2)}V_\extra^{-1}
\ee
where $M_F$ is the fundamental scale of gravity.
The graviton in these theories is a mode which is
homogeneous on ${\cal M}_\extra$. Contributions to the
gravitational potential due to excitations of the extra-dimensional
geometry (Kaluza-Klein (KK) modes) are Yukawa-like
and important only at distances short compared to the
inverse mass of the associated KK mode.

For $d=2$ and $3$, most compact manifolds
admit only a hyperbolic homogeneous geometry.
CHMs are obtained from their universal covering space $H^d$,
by ``modding-out'' by a (freely-acting) discrete subgroup $\Gamma$ of the
isometry group $SL(d,1)$ of $H^d$.
If the extra dimensions comprise a CHM \cite{KMST} then
all experimental constraints are met by $M_F\simeq TeV$ and a
maximum spatial extent of the CHM of only $\order(30) M_{F}^{-1}$.

An important property of CHM's is  the dependence of their volume on linear size:
$V \sim b_c^d e^{\beta}$ for a given curvature radius $b_c$,
with $\beta b_c$ proportional to the maximum linear dimension of the manifold.
Crucially, for a homogeneous geometry, $e^\beta$ is a
topological invariant of the manifold; in $d=2$, it is
proportional to the Euler characteristic. In
$d=2$ and $3$, there is a countable infinity of CHMs with volumes
distributed approximately uniformly from a finite minimum value
to infinity (in units of the curvature scale $b_c$).
In large volume CHM's there is
generically a gap in the spectrum of the Laplacian between the
zero mode (the graviton) and the next lowest mode. A theorem due
to Sarnak in $d=2$ and a conjecture due to Brooks in $d\geq3$
state (approximately) that for large volumes characteristically
$\mgap = {\cal O}(b_c^{-1})$.  The natural value for $b_c$
is ${\cal O}(M_F^{-1})$. One therefore expects
\be
M_{KK}\gsim M_F\ .
\ee
The solution to the gauge hierarchy problem
in this model is therefore at least technically natural,
corresponding to the combination of a large, but topological, value of $e^\beta$
and the natural value $b_cM_f={\cal O}(1)$.

Aside from its attractive particle-physics features,
this construction has important cosmological implications.
In the balance of this letter we show how the large scale homogeneity
and  local flatness of our universe emerge essentially automatically from
relatively generic local initial conditions and a specific choice for the topology of spacetime, and hence for $e^\beta$.

We imagine that the universe appears, in that its
geometry can first be treated classically, at a time $t_1$.
We assume that any initial evolution of the geometry of
$\M_\extra$ is brief and ignorable. This assumption is likely
justified if, as we assume, the mechanism stabilizing
the radius of curvature of $\M_\extra$  yields $b_c(t_1) = {\cal O}(M_F^{-1})$. The volume of $\M_\extra$ is
given by (\ref{Vextra}) with $e^{\beta} = {\cal O}\left(M_{Pl}^2/M_F^2\right)$.
We do not explain this large value of $e^\beta$, but note that it is
a topological invariant of $\M_{extra}$.

We assume further that $\M_{FRW}$ is expanding for $t\geq t_1$ with
massive KK modes
dominating the energy density of the universe.
We assume that these KK modes have initial number density of
$g_1 T_1^{3+d}$  ($g_1$
here being some slowly changing multiplicity factor, and $T_1\simeq M_F$ some effective temperature),
with energy density $g_1 T_1^{3+d} M_{KK}$.

As $\M_{FRW}$ expands, the KK modes are out of thermal
equilibrium and eventually decay, ultimately into light modes --
gravitons and standard model particles. We show below that
in this process most of the entropy in the universe moves from
the bulk to the brane.   We assume for simplicity that the
KK decay is instantaneous, occurring at
$t_2\equiv\tau_{KK} = M_{Pl}^2/(g_{KK} M_{KK}^3)$
($g_{KK}$ quantifies the multiplicity of KK decay channels).
With this approximation, we may equate the total energy content of the comoving
3-volume before and after the decay:
\be \label{coe}
e^\beta b_1^d a_1^3 g_1 T_1^{3+d} M_{KK} = g_2 T_2^4 a_2^3
\ee
where $a_{1,2}$ are the FRW scale factors at time $t_{1,2}$.
Assuming that the universe was effectively matter dominated from
$t_1 = [ 8\pi g_1 T_1^{4+d}/(3 M_F^{2+d})]^{-1/2}$ to
$t_2$, i.e.
$\left( a_2/a_1 \right) = \left( t_2/t_1 \right)^{2/3}$
we obtain

\begin{eqnarray}
\label{T2} T_2 &=&  \left( 3/(8 \pi ) \right)^{1/4}
g_{KK}^{1/2} g_2^{-1/4}  M_{Pl}^{-1/2} M_{KK}^{7/4} T_1^{-1/4}  .
\end{eqnarray}
If $T_2 \gsim 1 MeV$, which can easily be realized, nucleosynthesis is unaffected.
For example with $d=3$, $g_1 \simeq g_2 \simeq g_{KK} \sim 100$,
$M_{KK}=3 M_F$, $M_F=5$TeV, we find $T_2 > 1$MeV.
For a consistent cosmology, baryogenesis probably must occur during or prior to
the decay of the KK modes.

The decay of the KK modes leads to the usual radiation dominated era,
from $t_2$ to $t_3$, followed by a matter dominated era, from $t_3$ to today.

We next calculate the level of residual inhomogeneity we
expect in our universe today. We assume that at $t_1$,
when the initial distribution of KK modes is set, there
exists some correlation scale $\xi\simeq M_F^{-1}$, below which
fluctuations in all quantities  are highly
correlated, but above which all fluctuations are completely uncorrelated.
We assume further that on this scale
there are ${\cal O}(1)$ fluctuations about a well defined
mean density $\bar\rho={\cal O}(M_F^{4+d})$.
Most importantly, we assume the universality of the underlying distribution
from which the $\rho$ in each correlation volume is drawn.

In momentum space,
these fluctuations correspond to a gas of massive KK modes populating the bulk.
As $\M_{FRW}$ expands, the gas cools, and the KK modes decay via a
cascade into massless modes (i.e. $m\ll M_F$).
However, the massive KK modes cannot decay purely into gravitons
(or gravitinos) because
both KK modes and gravitons are eigenmodes of the
generalized wave operator for spin-2 particles on the CHM and
are therefore orthogonal on the CHM.  Furthermore, since
the graviton is the zero-mode of this operator on the CHM,
the product of the wavefunctions of any number of gravitons will still be homogeneous.
The decay rate for a massive KK state to pure gravitons
will involve an overlap integral between the KK wavefunction and the
multi-graviton wavefunction and will therefore vanish
at lowest order in perturbation theory.
This can also be understood in terms of a residual discrete isometry of the CHM,
inherited from  $H^d$, under which the KK modes
carry non-zero charges while gravitons are scalars.
(The need to prevent KK decay  to multi-gravitons may constrain deviations from
factorizability of $\M$ which break the discrete isometry.)

The brane, however, breaks the residual symmetry.  KK modes can decay into
Standard model particles which are confined to the brane, and thus are not
eigenmodes of the same operator.
The coupling of KK modes to Standard Model modes is typically unsuppressed,
compared to the coupling between KK modes,
except for the wave-function overlap suppression.
The KK modes therefore preferentially decay to Standard Model
particles propagating on the brane, or to these plus gravitons.
Since there are many light non-gravitational modes on the brane, and
only one massless gravitational mode in the bulk, we expect most
decays of KK modes to deposit their energy and entropy in
SM fields on the brane.

Entropy injection into the brane is a powerful homogenizing
processes in this cosmology. Consider a primordial
fluctuation $\delta \rho / \rho (\xi,t_1)$. So long as the
underlying distribution  of initial perturbations has
sufficiently compact support, the central limit theorem ensures
that the magnitude of this fluctuation at some length scale
$\lambda$ at some time $t$ is suppressed by a huge number ${\cal
O}\left(\sqrt{n}\right)$, where $n$ is the number of
appropriately  redshifted initial correlation volumes
($\xi^{3+d}$) which project into the 3-volume $\lambda^3$: \be n
\simeq \frac{ e^\beta b_c^d \lambda^3}{ \xi^{3+d} (a(t)/a_1)^3 }\
. \ee The growth of density fluctuations in time partially
balances this statistical averaging. Primordial fluctuations grow
as $t^{2/3}$ in the matter dominated era while in the radiation
dominated era are frozen \cite{peacock}. Thus, the density
fluctuations at some late time ($t>t_3$) at the scale $\lambda$
are related to the initial ones by:

$$
\label{drhobyrho_a}
 \frac{\delta \rho}{\rho} (\lambda ,t) \!
  \simeq
\frac{\xi^{(3+d)/2} (a(t)/a_1)^{3/2}}{e^{\beta
/2}b_c^{d/2}\lambda^{3/2}} \! \left(\frac{t_2}{t_1}\right)^{\!2/3}
\!\!\left(\frac{t}{t_3}\right)^{\!2/3} \!\! \frac{\delta
\rho}{\rho} (\xi,t_1) $$

At the time of last scattering $t_{ls}$, for the fluctuations at the apparent
horizon scale, $\lambda = \lambda_{ls}$, we have
\begin{eqnarray}
 &&\frac{\delta \rho}{\rho}  (\lambda_{ls},t_{ls})  \simeq 
2.6 \times g_1^{5/6}g_{KK}^{-11/12}g_2^{-3/8}   
M_{Pl}^{19/12} \xi^{(d+3)/2} \nonumber \\
&& \times T_1^{(20d+71)/24} T_3  
M_F^{-(d+2)/3} M_{KK}^{-19/8} T_4^{-5/2} \lambda_{ls}^{-3/2}
{\delta \rho \over \rho} (\xi,t_1)
\end{eqnarray}

\noindent Here $T_3=5.5\Omega_0 h^2$eV is the temperature at the beginning of
the matter dominated era, $T_4=0.3$eV is the temperature of the CMB at
last scattering and we take $\lambda_{ls} \sim 3 \times t_{ls}$, with
$t_{ls}=180000 (\Omega_0 h^2)^{-1/2}$yrs.
We assume $\Omega_0 = 1$, $h =0.7$.
It is easy  to see that the homogeneity of the
universe ($\delta\rho/\rho (3t_{ls}) \lsim 10^{-5}$) can be achieved for
reasonable values of parameters.
For example, to keep $T_2 >1$MeV we use the values of the
relevant parameters given below  eq. (\ref{T2}) for which we find
$\delta\rho/\rho \sim 4 \cdot 10^{-19}$ (assuming $\frac{\delta
\rho}{\rho} (\xi,t_1)$ of order one). Thus
homogeneity is achieved without fine tuning of the initial parameters.

Since the power spectrum of fluctuations behaves as
$\lambda^{-3/2}$, the scale at which
fluctuations become large (of order one) is safely small --- of
order $4 \times 10^{9}$m.

It is of interest to note that even prior to the homogenization
that occurs due to the decay of the KK modes and
the injection  of the associated entropy into the brane, an additional
homogenization mechanism is at work -- mixing due to the
non-integrability of flows in the CHM, i.e.  chaos.
CHMs are the paradigmatic arena for chaotic dynamics.
Ignoring the effects of gravitational feedback, these dynamics cause
gradients in the energy-density to be reduced as
$e^{-\kappa d}$,
where $\kappa$ is the Kolmogorov-Sinai (K-S) entropy \cite{KS} of the flow,
and $d$ is the distance the flow travels.  It has been shown
\cite{gott} that $\kappa\simeq V_{CHM}^{(-1/d)} \log{2}$.
Using $d\simeq \tau_{KK}\simeq M_{Pl}^2/M_{KK}^3$ and
$M_{KK}\simeq M_F$, we find
\[
\kappa d \simeq
\left( M_{Pl} / M_{F} \right)^{2(1-1/d)} \log2 / (M_F b_c)
\]
This is a huge factor, which suggests that by the time of KK decay,
at any given point in $\M_{FRW}$, the density is incredibly homogeneous across $\M_{extra}$.
Note that these flows are effective at eliminating only those gradients of the
density which  are in the extra-dimensional directions -- ordinary gradients are mostly
unaffected by this effect, unless $\M_{FRW}$ is also a CHM.
Nevertheless, this mechanism assures that the  full $\sqrt{n}$ suppression
of density fluctuations is realized.
(This homogenizing effect of flows has been pointed out in a cosmological context,
but with ordinary three-space as the CHM,
by other authors  looking for alternatives to inflation \cite{topocosmo},
or  at least alternative solutions to the homogeneity problem \cite{CSS}.)

What then of the flatness problem -- the fact that the universe today is
so very close to flat, with curvature radius
$a_c \simeq \frac{3000 h^{-1}Mpc}{\sqrt{1-\Omega_0}}$ ?
We must separate this problem into two pieces:\newline
a) The Global Flatness Problem: Is the manifold of three-dimensional space-time one with a topology that admits
a flat homogeneous metric, or one with a topology that admits a curved (spherical or hyperbolic)
homogeneous metric, or neither?
We do not solve this problem, but rather assume that the global topology of space time is consistent
with Euclidean geometry and a trivial topology (infinite extent in all directions).
This is at least a technically natural assumption, i.e. it is topological
invariant unchangeable by cosmological dynamics absent the formation of a singularity. \newline
b) The Local Flatness Problem: Given that the global topology is what it is,
why have fluctuations in the local curvature on cosmological scales not caused either
the recollapse of the visible universe, or its runaway expansion -- i.e. why
does the universe look so flat?

In the context of this scenario, the flatness problem is just the
homogeneity problem applied to gravity -- the effective
three-curvature in a particular domain in the FRW dimensions is
obtained by averaging over many independent domains in the bulk.
The curvature can be either positive or negative on a given
domain of radius $\xi$. If the distribution around zero curvature
is  symmetric, then by the central limit theorem, the mean
3-curvature today will be given by:

 \be \frac{1}{\bar{a}_4^2} =
\frac{1}{\sqrt n}\frac{1}{a_4^2} =  \frac{1}{\sqrt
n}\frac{1}{a_1^{2}}
 \left(\frac{a_1}{a_2}\right)^2
\left(\frac{a_2}{a_3}\right)^2 \left(\frac{a_3}{a_4}\right)^2 \
\ee
where $n$ is calculated at the scale of the visible universe,
i.e. the Hubble distance $cH_0^{-1}$.
Expressed in fundamental units, the initial energy density of the
universe is
\be
G_{4+d}\rho_1=\frac{g_1}{M_F^{d+2}} T_1^{4+d} \ ,
\ee
where $G_{4+d}$ is the fundamental (4+d dimensional)
gravitational constant and $g_1$ is the density of states. The
appropriately redshifted energy density today is
\be \label{ed}
G_{4+d}\rho_4 = \frac{g_1}{M_F^{d+2}} T_1^{4+d}
\left(\frac{a_1}{a_2}\right)^3 \left(\frac{a_2}{a_3}\right)^4
\left(\frac{a_3}{a_4}\right)^3 \ ,
\ee
where we have used
correspondingly redshifted powers in the matter and radiation
dominated eras.

The flatness problem is the smallness of the ratio between
$G_{4+d}\rho_4$ and $1/\bar{a}_4^{2}$; but now
\begin{eqnarray}
\frac{G_{d+4} \rho_{4}}{ 1/ \bar{a}^2_4} & \simeq &
\lambda^{3/2} M_{Pl}^{-7/12} M_F^{-(2d +4)/3}
M_{KK}^{-9/8}T_1^{(4d+37)/24}T_3 \nonumber \\
& \times & T_4^{5/2}a_1^2 \xi^{-(d+3)/2}
g_{KK}^{-1/12}g_1^{1/6} g_2^{7/8}
\end{eqnarray}
$T_3$ is, as before, the temperature at the beginning of the matter
dominated era, $T_4=3 \cdot 10^{-4}$eV is the temperature today while
$\lambda = cH_0^{-1} = 9.78 h^{-1} 10^9$lyrs. This
ratio is much greater than unity for any reasonable values of
parameters. To keep $T_2 \gsim 1$MeV, we use the values
of the relevant parameters as given below  (\ref{T2}) for which
$T_2 > 1$MeV giving $10^{8}$. Thus, the curvature term
today in Einstein's equation is negligible with respect to the
energy density term which achieves flatness without fine
tuning of the initial parameters.

The Kolmogorov-Sinai smoothing will play the same role as in the
homogeneity problem.

The net effect of these mechanisms is to solve  the local
aspect of the entropy problem. 
The entropy within the horizon at some earlier time (say
CMBR) was much smaller than the entropy within the horizon today,
thus implying that the horizon today consists of many causally
disconnected regions, and yet the universe appears uniform. 
In the context of our homogenization mechanism  (and its assumptions about the 
topology of $M_{\extra}$), this is no longer a problem.
We should note that, since we assumed global Euclidean geometry and trivial
topology of the 3-space, the global entropy problem (the existence of the huge
dimensionless number of $10^{88}$, as discussed in \cite{Linde}) is
equivalent to the global flatness problem --- why the universe is so big
(infinite in extent). As we stated earlier, we do not solve this problem.

We have thus shown that, under certain assumptions, the homogeneity and
local flatness of the visible universe
is a generic feature of a universe with large extra dimensions
and factorizable geometry. To do so
we have had to assume that the underlying distribution functions from which density and
curvature are drawn are homogeneous.  Additionally, the scenario
places certain demands on the topology of space-time -- it
must be factorizable $\M=\M_{FRW}\times\M_{extra}$,
with $\M_{FRW}$ homeomorphic to $R\times E^3$, and the topology of $\M_{extra}$,
as quantified by $e^\beta$, sufficiently complex.  Moreover, it is apparent
that the topology of $\M_{extra}$ must not vary,
or at least not  by a lot, from point to point in $\M_{FRW}$.
We make no attempt here to quantify the limits on how
much the topology may vary, for it clearly depends on whether
the only effect will be a change in the volume, and hence the
value of the effective gravitational coupling $G_N$.
In that case the limits are likely quite weak since we know
only that $\Delta G_N/G_N \lsim 0.1$ over the relevant length and
time scales.  However, if other constants of nature, such as
fermion masses, or $\alpha_{EM}$ are very sensitive to changes
in the topology, then more restrictive limits may apply.

We have not asserted any solution for the origin of structure --
though we may hope that fluctuations of the brane might be the source --
nor have we addressed the issue of baryogenesis -- though we note
that the decay of the KK modes is  out of equilibrium and could
be baryon-number and CP violating.  Further, we acknowledge that there
are difficulties for proton stability, though these have been previously
discussed in the context of the canonical scenario \cite{ADD}.

In summary, while we do not claim at present that this is a complete
alternative view of a universe with neither fine-tunings nor inflation,
we have demonstrated some interesting cosmological consequences of large
extra dimensions, warranting further detailed study.

This work was supported by NSF grant PHY-0094122 (MT) and by the DOE 
(GDS and DS).

\end{document}